# The Policy Deficit in AI x Social-Emotional Learning Research

Tran Van Cuong[1]*, Liu Yihan[2], Nguyen Van Tuong[3, 4]

[1] Department of Computer Science, University of Copenhagen, Sigurdsgade 41, Copenhagen, 2200, Denmark. Email: trancuongk17@gmail.com. ORCID ID: 0000-0001-5766-5802

[2] Doctoral School of Education, University of Szeged, Petőfi Sándor sgt. 30–34, Szeged, H-6722, Hungary. Email: yihanliu550@gmail.com. ORCID: 0009-0004-9186-2809

[3] University of Social Sciences and Humanities, Vietnam National University, Ho Chi Minh City, Quarter 6, Linh Trung Ward, Thu Duc City, 71308, Vietnam. Email: tuongnguyen@hcmussh.edu.vn. ORCID ID: 0000-0001-8951-4697

[4] Vietnam National University, Ho Chi Minh City, Quarter 6, Linh Trung Ward, Thu Duc City, 71308, Vietnam.

**Abstract**: As artificial intelligence (AI) is increasingly integrated into social-emotional learning (SEL) initiatives, the need for evidence-based policy has become paramount. We systematically reviewed 65 peer-reviewed papers that examine the intersection of AI and SEL to investigate how these studies articulate policy implications. Our analysis revealed a substantial "policy deficit" in the current AI × SEL literature: nearly three-quarters of the studies did not mention policy implications at all. Using the "WH-question" framework (Who, What, Why, When/Where, and How), we map the policy implications narratives present in the literature and show that they often lack the specificity and actor-oriented guidance required for effective evidence-informed policymaking. We find a significant association between publication venue and policy engagement, suggesting that current academic incentive structures may prioritize technical innovation and pedagogical feasibility over explicit engagement with governance and regulation. This study identifies a "techno-solutionist" trap, where technical potential is foregrounded while the institutional conditions for responsible implementation remain under-specified. We conclude by proposing a shift from "implication-as-afterthought" to "implication-as-methodology" and offer a set of actionable guidelines for researchers, editors, reviewers, and policymakers to bridge the gap between AI innovation and educational governance. Rather than presenting policy as a generic ethical horizon, we argue that AI-SEL studies should systematically specify Who should act, What actions are recommended, Why these actions are needed, When and Where they apply, and How strongly they are framed, thereby strengthening the translation of AI × SEL innovation into educational policy and practice.

**Keywords**: AI × SEL, Artificial Intelligence, Social-Emotional Learning, Policy Implications, Evidence-Based Policy, Systematic Review.

**Impact Statement**: Artificial Intelligence (AI) and Social-Emotional Learning (SEL) are at an intricate intersection with a growing influence on society. However, our review of 65 research studies shows a major gap: while many papers praise the potential of AI for SEL, only one-fourth provide policy recommendations. Few offer detailed, actor-specific recommendations on policy issues such as privacy, teacher training, and resource allocation. We identify this "policy deficit" and introduce a new writing framework that prompts researchers to provide clear, actionable guidance. We are calling for a culture shift in science to ensure that research is consistently coupled with practical policy recommendations.

## Introduction

Over the past decade, artificial intelligence (AI) has become increasingly entangled with social and emotional learning (SEL) research and adjacent domains, such as emotional assessment, soft skills, and socio-emotional development in education. Across systematic reviews, mapping reviews, and conceptual papers, AI is often presented as a powerful tool (including but not limited to emotion-recognition systems, chatbots, and large language models (LLM), LLM-based tutors, and AI-enhanced toys) that can scaffold students' self-awareness, emotional regulation, social skills, and broader socio-emotional competencies (Guilbaud et al., 2022; Liu et al., 2026; Özdemir Beceren et al., 2025; Sethi & Jain, 2024). This "innovation-centric" research highlights a compelling trend in the field and opportunities for more engaging, adaptive, and inclusive social-emotional teaching and learning, particularly when AI systems provide real-time emotional feedback, simulate social interactions, or tailor prompts and scenarios to individual learners (Henriksen et al., 2025; Liu et al., 2026).

Yet, this technical expansion has significantly outpaced the field's capacity to govern it. A parallel "governance-centric" body of literature on emotional AI, children's rights, and AI in education in general has raised critical concerns about surveillance, bias, depersonalization, and the commercialization of learners' emotional data, especially in relation to marginalized and other vulnerable groups. McStay (2020), for instance, critically examined facial-coding emotional AI used to quantify SEL, arguing that such technologies raise serious questions about validity, child rights, and the clash between commercial interests and the public good in education. Vistorte et al. (2024) systematically reviewed AI-driven emotion assessment in learning environments and identified challenges around accuracy, privacy, and cross-cultural validity. Similarly, Atabey & Scarff (2023) argued that emotional AI in education contradicts the General Data Protection Regulation's (GDPR) fairness principle and represents a potential threat to children's rights to freedom of thought, non-discrimination, privacy, and data protection.

To move beyond this dichotomy, this systematic review contends that policy should not be treated as a peripheral horizon at the end of a study, but as a core research objective. To facilitate this, we adopt a novel methodological lens: the "WH-question" framework (Who, What, Why, When/Where, and How). By mapping the implications reported in 65 peer-reviewed papers against these five interrogatives, we identify not just the absence of policy implications, but the various aspects of policy implications that are missing.

We argue that the current literature requires a substantial reorientation: from "implication-as-afterthought" to "implication-as-methodology". This approach ensures that researchers do not merely list high-level ethical concerns but provide the specific, actionable instruments that policymakers and educators need to navigate the intersection of AI and SEL. The following sections evaluate the current landscape of AI × SEL research through this framework, highlighting the structural barriers to evidence translation and offering concrete guidelines to bridge the gap between technical potential and institutional reality.

**AI × SEL reviews and policy implications**

To investigate whether evidence-synthesizing literature offers evidence-grounded policy guidance, we first examined whether evidence-grounded policy guidance is offered in some systematic reviews.

We first look at a systematic mapping review on AI-powered SEL in teacher education by Méndez et al. (2026), which identifies key trends, such as AI-enabled personalized SEL training and predictive engagement analytics, and documents institutional resistance and insufficient structured training as obstacles to SEL implementation. Their conclusion explicitly calls for context-sensitive, ethically grounded teacher education programs and proposes directions for policy development. While suggesting that policymakers play an important role in AI-powered SEL, they did not directly suggest any actionable policy recommendations for them.

Henriksen et al., (2025) provide a critical literature review of generative AI and SEL, introducing a conceptual framework to prepare pre-service and in-service teachers to navigate risks such as depersonalization, bias, and privacy violations. The framework emphasizes ethical considerations, human oversight, and cultural sensitivity, effectively embedding policy implications into teacher education design: teacher programs are positioned as key sites for operationalizing ethical and cultural safeguards in AI-augmented SEL. However, the framework is only "a starting point in policy discussions". Though the authors suggested several policy guidelines, they did not specify the sources for these guidelines. This limits the trustworthiness and clarity of the guidelines.

In another review, Liu et al. (2026) map LLM chatbots against 15 SEL competencies and 19 affordances, identify five ethical risks (transparency, privacy, equality, beneficence, and affect/identity safety), and explicitly state that policymakers should prioritize ethical guidelines and support specialized, equitable LLM tools for SEL. However, the authors did not specify any policy recommendations or implications for policymakers. Another review by Sethi & Jain (2024) on AI technologies for SEL suffers from the same issue: no explicit policy guidance was offered.

To conclude, though the field has several systematic reviews to synthesize evidence at a high level, very few papers provide detailed, evidence-based policy implications that specify responsible actors, contextual conditions, and concrete courses of action. Rather, these reviews only mentioned policy at a high level by calling for ethical guidelines, equitable access, or robust ecosystems without pinning down specific instruments or actor responsibilities (Liu et al., 2026; Sethi & Jain, 2024; Vistorte et al., 2024). Some other works particularly engage with policy frameworks, which we consider a form of theory, while insufficiently engaging with policymakers as a potential reader to uptake their findings (Atabey & Scarff, 2023; Xiao & Gonçalves, 2025). In these works, policy language functions as a broad horizon for future work rather than a clearly specified object of analysis that provides useful information for policymakers.

## Current study and guiding framework

### *Conceptual Foundation*

The present study adopts the definition of social emotional learning (SEL) proposed by Collaborative for Academic, Social, and Emotional Learning (CASEL, 2024), which conceptualizes SEL as

> "an integral part of education and human development. SEL is the process through which all young people and adults acquire and apply the knowledge, skills, and attitudes to develop healthy identities, manage emotions and achieve personal and collective goals, feel and show empathy for others, establish and maintain supportive relationships, and make responsible and caring decisions."

This systematic review positions itself at the intersection of AI × SEL. This intersection can be provisionally defined as the emerging domain through two primary strands: the integration of SEL constructs into the architecture of AI models (*SEL inside AI*), and the utilization of these systems to assess, support, or influence learners' competencies (*AI for SEL*). The intersection may include, but is not limited to AI-supported interventions, affective-computing technologies, and predictive analytics for SEL-related outcomes. Together, these strands introduce complex ethical, pedagogical, and governance challenges. Addressing these challenges necessitates an emerging policy domain for AI × SEL. Particularly, instead of treating "policy implications" as a peripheral paragraph at the end of empirical or review articles, we treat them as the primary object of inquiry. By systematically identifying, coding, and interpreting the implications reported in AI × SEL research, including those framed as legal, regulatory, institutional, professional, or ethical, the review seeks to clarify how policy is imagined in this emergent field and to what extent these imaginations provide specific, feasible, and context-sensitive guidance for decision makers and practitioners working at the interface of AI × SEL. To this end, policy is conceptually defined as "governance principles that guide courses of action and behavior in organizations and societies" (Aguinis et al., 2022).

### *Theoretical orientation*

Drawing on the concepts of evidence-based policymaking (EBPM) and evidence-to-policy translation (Ingold & Monaghan, 2016), this study treats policy narratives in research papers as potential knowledge brokers between research communities and educational policy actors (Chalmers, 2005; Oliver et al., 2014). We examine whether and how they fulfil this brokering role by articulating explicit policy implications. EBPM and policy translation depend on a chain of aligned processes, actors, and events that collectively enable policy development (Ingold & Monaghan, 2016). Within this chain, explicit and well-structured policy recommendations are not merely communicative devices but functional links that facilitate the movement of evidence into decision-making arenas (Derstroff et al., 2026; Dwivedi et al., 2024). Their presence signals an intentional effort to bridge the epistemic and institutional boundaries that often separate research from policy practice. Moreover, presenting findings in formats that directly target potential users—in this case, policymakers—constitutes a critical mechanism for promoting the uptake and use of research (Nutley et al., 2007). In this sense, the articulation of policy implications within evidence syntheses can be read as both a communicative act and a structural condition for evidence-informed governance.

*Analytical framework*

Building on this theoretical foundation, we used the "WH-question" framework from Dello Russo et al. (2023) to organize and interpret the policy implications found in the corpus. In this approach, we look at each policy implication as a short narrative that can be broken down into several parts: who it speaks to (for example, ministries, quality-assurance agencies, university leaders, program directors, instructors, or students), what kinds of actions it suggests, why those actions are considered important (the expected outcomes or possible trade-offs), when and where they are meant to apply (specific contexts, sectors, regions, or stages of AI adoption), and how the statements are framed — whether as obligations or possibilities, using words like *must*, *should*, *could*, or *is needed*. This way of coding the literature helps us see not only what kinds of policy advice are being offered but also how authors position their recommendations and the level of certainty or authority they imply.

In this research, we ask the following five questions in the context of AI × SEL research:

1. Why are these actions considered necessary or valuable, i.e., what outcomes, rationales, or trade-offs are mentioned?
2. What types of policy actions or recommendations are proposed?
3. Who are the main policy actors addressed in AI-SEL research?
4. When and where are these policy implications situated, i.e., which educational levels, sectors, regions, or stages of education?
5. How are these implications framed linguistically?

## Methods

### Literature search

We adapted the PRISMA guideline (Page et al., 2021) when conducting and reporting this review. We developed the search string based on two core concepts: "social emotional learning" AND "Artificial Intelligence". To capture the breadth of the field, we identified relevant synonyms and related terms for each concept and combined them using Boolean operators. We searched Scopus and Web of Science because these databases provide broad coverage of peer-reviewed research across education, psychology, computer science, and related disciplines. Full search strings are provided in the Supplementary File. We searched among titles, abstracts, and keywords, with no time frame limitation.

The search yielded 414 records. After removing 122 duplicates, we retained 292 records for further screening. We then excluded erratum notes, non-English records, and one retraction publication, leaving 271 records for eligibility assessment.

**Eligibility criteria and study selection**

Screening was conducted in two stages. First, titles and abstracts were reviewed against the eligibility criteria. Records that did not meet the inclusion criteria were excluded. The remaining studies were then subjected to full-text assessment. Articles were excluded at this stage when they did not meet the criteria. The inclusion and exclusion criteria are:

1. Empirical studies based on primary or secondary data, i.e., meta-analyses and systematic reviews are included. Non-empirical works (e.g., conceptual papers or commentaries) are excluded.
2. Focus on the intersection of artificial intelligence (AI) and social-emotional learning (SEL) at any stage of education, from preschool to higher education and adult learning. Studies about educational technology without a focus on AI are excluded.
3. Peer-reviewed papers in journals or proceedings are included. Non-peer review papers, reports, or working papers are excluded.

Following title and abstract screening and full-text review, 65 articles were retained for analysis. The process resulted in 65 articles in the final sample for analysis (Figure 1).

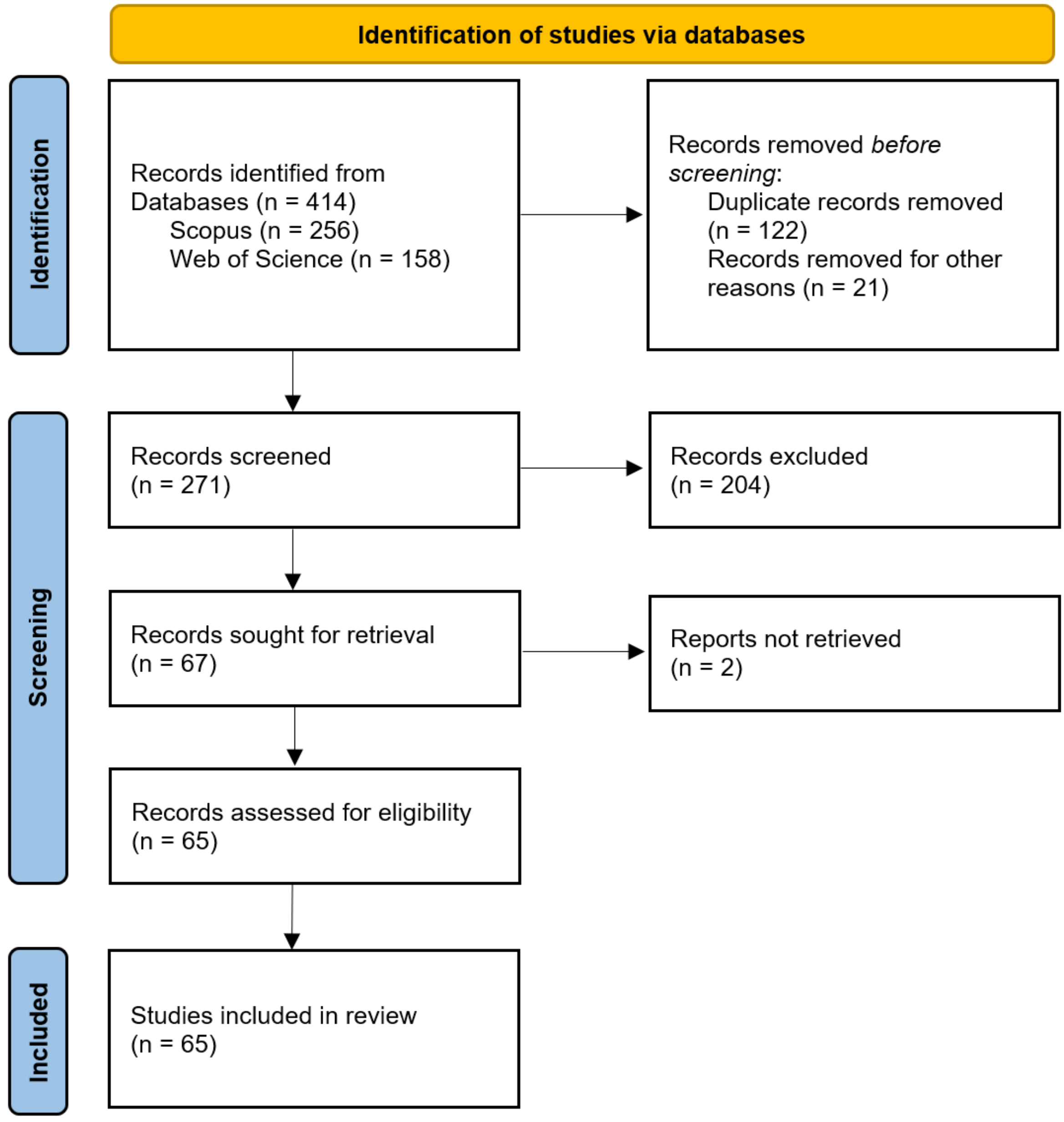

Figure 1. PRISMA flow of the current review

**Coding procedure**

Our analysis focused on how policy implications are articulated in AI–SEL research. After completing the selection and screening procedures aforementioned (step 1), we compiled all implications from the included studies into a single corpus (step 2). The characteristics of studies are presented in the Supplementary File to enhance the transparency and reproducibility of our analysis. Descriptive results are aggregated at the paper level. But for analysis with the "WH-question" framework, we treated each distinct implication statement as a unit of analysis.

We defined policy implications as logical consequences of the study findings for policy. They can be operationalized as explicit or implicit implications. Explicit policy implications are written with

keywords specifically related to policy, including policies and policymakers. For example, Pino Tarragó et al. (2025) stated that "institutional policies must be oriented toward innovation, interdisciplinarity, and continual renewal". We identified 15 papers with explicit policy implications. Implicit policy implications do not have explicit keywords above, but they target educational leaders or educational organizations. In these cases, policies can be derived from the sentence by readers. For example, Yana-Salluca (2025) stated that "it is recommended that higher education institutions consider the implementation of AI-based technologies as part of their pedagogical strategies". This statement does not explicitly mention policy implications, but it still directs a course of action that governs schools. It implies that higher education institutions may need policies to support the integration of AI into teaching and learning. Coding implicit policy implications can be subjective; thus, two authors conducted an individual coding process and compared the results and reached agreement in discussion with a third author. As a result, we identify three papers containing implicit policy implications.

To organize and interpret these implication statements, we adopted the "WH-question" framework proposed by Dello Russo et al. (2023), which conceptualizes implications as short narratives that can be decomposed into a set of core interrogatives (who, what, why, when/where, how). In applying this framework, we first extracted every sentence in which authors addressed policymakers, institutions, or other organizational actors, or used implication-like language (e.g., "policy makers should", "institutions need to", "there is a need for regulations"). Each of these claims was treated as an "implication narrative". For each narrative, we coded the following dimensions:

- Why – the rationales to justify the suggestions of implications (e.g., protecting children's rights and privacy, ensuring fairness and non-discrimination, improving SEL outcomes, managing risks of surveillance or depersonalization, addressing digital inequities).
- What – the types of policy actions or recommendations proposed, such as developing or revising laws and regulations, issuing guidelines or standards, mandating training, investing in infrastructure, establishing oversight mechanisms, or restricting certain forms of AI use.
- Who – the primary policy actors explicitly or implicitly addressed (e.g., ministries or government agencies, quality-assurance and accreditation bodies, university or school leaders, program directors, instructors/teacher educators, students, technology vendors and developers).
- When and where – the situational anchoring of each implication (e.g., educational levels (early childhood, K-12, higher education), sectors (public vs. private), geographic regions, and stages of AI adoption (pilot experimentation, scaling, or consolidation) to which the implications were said to apply.
- How – the linguistic and normative framing, focusing on modal verbs and stance markers (e.g., must, should, need to, could, is likely to) to capture the implied strength of obligation, certainty, or tentativeness in each implication, and the use of policy implication markers within a paper to signify the presence of implications.

The five research questions of this review were directly mapped onto these analytic dimensions. Coding was guided by an a priori codebook derived from the WH-question framework and iteratively refined during pilot coding of a subset of studies. When a single narrative contained multiple elements (e.g., addressed more than one actor or proposed several actions), we allowed multiple codes within a dimension but retained the narrative as a single unit to preserve its coherence. Disagreements in coding were resolved through discussion and, where necessary, by revisiting the original full-text context to clarify the intended audience, scope, or rationale.

By structuring the analysis around these WH-questions, we move beyond simply listing policy recommendations and instead synthesize implications for policy.

## Findings

### Descriptive findings

Despite the volume of research (65 included papers), there is a notable lag in translating findings into concrete policy recommendations. Regarding policy implications, our analysis reveals that approximately 72% of the sampled literature ($n$ = 47) did not articulate any substantive consequences for policy governance. 28% ($n$ = 18) provide implications for policy, among them, three articles offer implicit cues. Among these 18 papers, one contains four policy implications, two provide three, six offer two, and the remaining nine contain only one policy claim each. Taken together, these descriptive findings indicate that policy implications are present in only a minority of AI × SEL studies and, when they do appear, are relatively limited in the number of policy-relevant claims.

Among papers that contain policy implications, only two studies were recorded during the 2021-2023 period, focusing on robotic toys and general reviews. Three studies were published in 2024. During 2025-2026, a total of 13 studies were concentrated, reflecting the recent explosion in AI and its application in educational settings. Research methods range from small-scale qualitative interviews to large-scale quantitative surveys. Notably, as the field has grown into 2026, we see a trend toward larger sample sizes (e.g., surveys with 1172 undergraduate students or 537 English teachers), moving beyond initial exploratory case studies.

A Mann-Whitney U test was conducted to determine if the presence of explicit policy implications in the reviewed papers was associated with differences in sample size among 45 papers that reported human sample sizes. The analysis revealed that papers containing policy implications ($Mdn$ = 27.42, $n$ = 12) did not significantly have a higher sample size than those without such implications ($Mdn$ = 21.39, $n$ = 33), $U$ = 145.00, $z$ = –1.36, $p$ = .173.

These data suggest a structural incentive problem in the field. Statistical analysis indicated that publication venue significantly dictates policy engagement, $\chi^2(1, 65) = 15.02, p < .001$. Papers published in journals were significantly more likely to contain policy implications ($n$ = 16, 45.7%) compared to those published in conference proceedings ($n$ = 1, 3.3%). While our data do not allow us to infer causality, this pattern suggests that some publication venues may prioritize technical novelty over explicit engagement with policy and governance.

A Fisher-Freeman-Halton exact test, computed via Monte Carlo simulation (B = 20,000), was performed to examine the association between study type and the inclusion of explicit policy implications. The analysis indicated a statistically significant association, $p = .015$. Qualitative research ($n$ = 6, 85.7%), survey ($n$ = 4, 80.0%), and reviews ($n$ = 5, 35.7%) demonstrated a higher proportionality of policy implications compared to development, evaluation, and experimental studies ($n$ = 2, 5.1%).

The same test indicated a statistically significant association between educational level and the presence of policy implications, $p = .041$. Descriptive analysis revealed varying proportions of policy implications across educational levels: Higher Education (40.0%), K-12 (32%), adult education (including SEL at work and among adults without specific levels of education; 25.0%), and early childhood education (20.0%). In contrast, there are three studies involving mixed-level populations, none of which demonstrated policy implications.

While these statistics may reveal a structural issue in academic publishing, they do not capture the substance of the few policy implications that do exist. To understand the quality and utility of these implications, we now turn to our qualitative analysis using the WH-question framework. We have identified 31 policy implications claims. In the sections below, these claims serve as the unit of analysis.

**The Why**

The analysis of the implications reveals that the proposed policy implications are not merely speculative, but rather driven by specific structural, ethical, and instructional necessities. The rationales undergirding the proposed implications are organized into three primary dimensions

*Safeguarding Vulnerable Populations and Mitigating Risk*

The primary rationale for stringent and immediate policy intervention is the critical need to protect young and vulnerable learners within AI-mediated environments. Current empirical applications call for proactive regulatory measures to safeguard the privacy, data integrity, and developmental well-being of young learners (Aldhilan & Rafiq, 2025; Samawi & Al-Assaf, 2023). Because emotional AI interactions carry inherent psychological risks, formal guidelines are required to mitigate specific risks within vulnerable populations and establish evidence-based boundaries that protect privacy without stifling meaningful and supportive interactions (Pnevmatikos & Christodoulou, 2026; Xiao & Gonçalves, 2025). Furthermore, institutional mandates are proposed to build public trust and ensure robust legal compliance when handling sensitive student data (Aure & Cuenca, 2024; Pino Tarragó et al., 2025).

### *Institutional Efficacy, Structural Alignment, and Future Readiness*

From an organizational standpoint, implications regarding infrastructure and resource allocation are proposed to overcome the limitations of generic technological tools, heading toward AI tools for customized learning experiences (Deng & Ouyang, 2026). Establishing specialized frameworks is necessary to solve the quality gap inherent in using open-domain, generic AI models for specialized education (Xu et al., 2026). Strategic policies aim to:

- Bridge socio-economic gaps to ensure universal access and prevent systemic technology divides (Samawi & Al-Assaf, 2023).
- Prevent exclusionary practices by developing adaptive, universal pedagogical toolboxes (Kewalramani et al., 2021).
- Ensure resources are technically, pedagogically, and psychologically sound through interdisciplinary collaboration (Tsai & Wu, 2026).

Ultimately, these mandates seek to drive leadership-led, purposeful scaling within ethical boundaries, transforming AI from an isolated, standalone tool into an integrated core component of classroom instruction (Pino Tarragó et al., 2025).

### *Preserving Human Agency and Calibrating the Pedagogical Climate*

At the instructional level, implications are proposed to manage the evolving socio-technical ecosystem of the classroom. As technology alters traditional instructional dynamics, authors advocate for developing guidelines to provide a clear roadmap for educators navigating their shifting roles (Tsai & Wu, 2026) while preparing pre-service teachers for inevitable systemic shifts (Xu et al., 2026). Crucially, these frameworks should be designed to prevent the dehumanization of education by balancing technological optimization with human mentorship (Deng & Ouyang, 2026) and learners' readiness (Xiong, 2026). In terms of student-facing interactions, affective guidelines are proposed to ensure that AI-driven experiences do not leave students emotionally overwhelmed by negative narrative elements, instead intentionally cultivating student resilience, critical thinking, and core human socio-emotional competencies through balanced, supportive emotional design (Jafari et al., 2026).

## The What

Our analysis reveals three themes of policy implications being proposed, including institutional mandates, resource allocation, and ethical frameworks.

### *Institutional Mandates and Governance*

The strongest theme is the need for institutional mandates and governance. Several implications call for formalized policy structures such as data governance (Aldhilan & Rafiq, 2025; Fu & Wu, 2026), institutional innovation (Pino Tarragó et al., 2025; Samawi & Al-Assaf, 2023), adaptive governance (Aure & Cuenca, 2024), global alignment (Pino Tarragó et al., 2025), and collaborative governance (Samawi & Al-Assaf, 2023) to ensure AI-supported SEL is implemented consistently rather than left to individual discretion. Related recommendations emphasize interdisciplinary co-design (Tsai & Wu, 2026), ecosystem synergy (Xu et al., 2026), and stakeholder collaboration (Aure & Cuenca, 2024; Henriksen et al., 2025; Pnevmatikos & Christodoulou, 2026; Su et al., 2024; Xiao & Gonçalves, 2025), indicating that policy should coordinate schools, teachers, researchers, and other actors.

*Resource Allocation and Strategic Infrastructure*

Several implications emphasized flexible resource allocation (Xiong, 2026), institutional capacity building for AI literacy, domain-specific alignment (Xu et al., 2026), instructional scaffolding, teacher mentorship, optimized AI adoption (Deng & Ouyang, 2026), and teacher training (Aldhilan & Rafiq, 2025; Tsai & Wu, 2026) as necessary conditions for implementation. In practical terms, this means policy should fund training, infrastructure, and implementation support rather than assuming schools can absorb AI-related change without additional investment.

*Ethical Frameworks*

Multiple implications highlight regulatory standards for ethics and equity (Henriksen et al., 2025), data governance (Aldhilan & Rafiq, 2025; Fu & Wu, 2026), child-centric data protection (Samawi & Al-Assaf, 2023; Xiao & Gonçalves, 2025), transparency (Aldhilan & Rafiq, 2025), safety protocols (Pnevmatikos & Christodoulou, 2026), and ethical adoption (Fu & Wu, 2026). Together, these implications frame AI × SEL as a governance issue as much as a pedagogical one, with strong attention to rights, inclusion, and harm prevention. However, these ethical frameworks, while necessary, often remain at the level of high-level suggestions rather than specific implementation mechanisms, reinforcing the tendency to view policy as an abstract concept.

**The Who**

Policymakers are explicitly mentioned 15 times (48%), while educational authorities are mentioned seven times (via words such as educational institutions or educational leaders). In four cases, it is unclear who the actors of those implications are. 12 claims mix those policy actors with other stakeholders such as educators and curriculum designers, making it difficult for us to disentangle what the policy implications are and how the policy implications being made differ from other implications.

**The Where and When**

Overall, very few papers have specified the condition for when and where the policy implications are applicable. The lack of conditional specificity is particularly striking; researchers frequently offer broad, generalized recommendations that fail to clarify the precise geographic, sectoral, or developmental contexts to which they apply. Without this situational grounding, many policy suggestions end up being difficult to act on. They are disconnected from the varied realities of educational settings and the timing constraints that shape implementation.

To overcome this disconnect, policy implications must be grounded in the specific realities of educational ecosystems. We can resolve this lack of situational grounding by categorizing these recommendations across three distinct levels of education: the macro-level, which addresses systemic foundations and governance; the meso-level, which focuses on institutional integration and professional infrastructure at the school level; and the micro-level, which targets the immediate pedagogical practices and student learning and well-being concerns of the classroom. Below, we thematically present our findings, which reveal that these educational levels are only fragmentarily addressed in the literature, while offering minimal insights into the question of when, indicating the lack of specificity.

*Macro-Level: Systemic Foundations and Governance*

16 implications target the macro-level of policy. Macro-level implications prescribe policy recommendations for AI in education in general terms. They suggest that policymakers develop comprehensive policy frameworks taking into account systemic equity (Kewalramani et al., 2021; Liu et al., 2026), child-centric safety (Aldhilan & Rafiq, 2025; Samawi & Al-Assaf, 2023; Xiao & Gonçalves, 2025), privacy (Henriksen et al., 2025), technology optimization (Xu et al., 2026), and encourage interdisciplinary collaboration (Tsai & Wu, 2026). As for the question of when, we found that policymakers must act proactively or even immediately (Xiao & Gonçalves, 2025) to create the regulatory environment necessary for safety. We also include in this category suggestions targeting "all kindergartens, irrespective of socio-economic background" (Samawi & Al-Assaf, 2023) as well as "local and international privacy laws" (Aldhilan & Rafiq, 2025).

*Meso-Level: Institutional Integration and Professional Development*

14 claims out of 31 claims are implications at the meso level. Meso-level implications emphasize that technology success depends on the school's internal culture and infrastructure. These implications target educators, policymakers, and educational authorities or leaders. The implementation here is continuous, reflecting the need for iterative refinement of policy (Aure & Cuenca, 2024; Pino Tarragó et al., 2025). These implications suggest that institutions must move away from static rulebooks, instead adopting a comprehensive, "living" AI policy grounded in the belief that clear guidance can help mitigate concerns (Aure & Cuenca, 2024), and being flexible in resource allocation (Xiong, 2026). This involves school comprehensive guidelines (Pnevmatikos & Christodoulou, 2026) or institutional-level "multi-layered ecosystems" (Xu et al., 2026) that support teacher training (Fu & Wu, 2026; Xu et al., 2026), ongoing meta-reflection of teachers on using AI (Aldhilan & Rafiq, 2025; Aure & Cuenca, 2024), and the adaptation of learner-centric AI tools (Deng & Ouyang, 2026).

*Micro-Level: Pedagogical Practice and Student Wellbeing*

Only one claim targets policy at the micro-level, focusing on the moment of interaction within classrooms. Jafari et al., 2026 stated that "accompanying teacher guidelines should include age-appropriate strategies for emotional regulation and constructive response". The primary responsibility here lies with educators to use AI as a catalyst for resilience rather than a source of stress. To be more specific, the guideline may advise teachers to "balance negative emotions with coping strategies", ensuring that when AI-delivered content evokes fear or sadness, the educator facilitates a constructive response, such as discussing "how communities rebuild after tragedy" (Jafari et al., 2026).

**The How**

Finally, we investigated how policy implications are rhetorically framed and how they are written between sections.

*Rhetorical Language Within the Claims*

We found that the largest group of rhetorical styles is prescriptive (24 claims), issuing clear directives for institutional and systemic action. This language is rigid, definitive, and uses authoritative verbs such as "must", "require", or "need to". Some other phrases categorize this group are "demand immediate attention", "it is necessary", and "becomes essential", which also, to some extent, convey the sentiment of urgency. Some illustrative quotes are: "Educators and policymakers must collaborate to ensure that this platform aligns with teaching goals and standards" (Su et al., 2024) and "policymakers need to consider comprehensive policy frameworks that address ethical considerations" (Aldhilan & Rafiq, 2025).

Another group of languages is tentative (seven claims). It utilizes verbs like "could", "may", "can", and phrases like "can be proposed" or "is recommended" to offer policy recommendations. While still authoritative, it functions as expert counsel meant to steer institutional administrators toward optimal frameworks rather than establishing legal or ethical absolutes. Some illustrative quotes are: "It is recommended that higher education institutions consider the implementation of AI-based technologies as part of their pedagogical strategies" (Yana-Salluca, 2025) and "For higher education institutions to successfully impart AI skills to students, they may need to redefine their role" (Aure & Cuenca, 2024).

*Rhetorical Language Within the Paper*

Interestingly, we found no section containing the terms implications for policy or policy implications. In the majority of the cases, implications claims appear under the heading of discussion or conclusion without any specific words signifying the existence of implications (*n* = 19). Five claims appear under the heading of "practical implications", two under the heading of "recommendations", while one is under the heading of "conclusions, implications, and provocations". Only two abstracts (3.1%) claim that the paper has implications for policy. These indicate that within AI × SEL research, policy implications are neither prioritized nor positioned as a focal point of impact.

We noticed some papers (out of our final 18 papers) provide a misleading or incomplete promise by stating to have implications for policy in the required impact statements or introductory text, yet proceed to ignore them in the main text. We decided not to cite these studies as they provide no implications for policy, yet we reported these as concerning observations of how researchers are shallowly engaging with policy implications to fulfill the requirements outlined by the editors and reviewers.

## 4. Discussion

The findings of this systematic review suggest a significant "policy deficit" in the AI × SEL literature. While there is a surge in empirical studies exploring AI's technical potential for SEL (65 papers), there is a persistent gap in how these findings are translated into evidence-informed policy guidance. Our analysis suggests that the vast majority of studies remain confined to technical efficacy or pedagogical feasibility, essentially functioning as "innovation-first" narratives and concerns with practical implications that are not policy.

### The "Techno-Solutionist" Trap

Our review reveals a field currently dominated by "techno-solutionism": the belief that AI applications will inherently solve SEL challenges, without sufficient engagement with the institutional or regulatory hurdles that prevent such solutions from scaling. The scarcity of policy implications suggests that researchers often position AI tools as viable solutions, overlooking the fact that SEL integration is largely an act of governance. When policy is mentioned, it is often broad, abstract, and lacking in specific, actionable instruments, fundamentally treating policy as a "horizon" rather than a set of practice-oriented tools. This disconnect risks creating a "pilot-study paradox" where robust experimental results fail to achieve real-world impact because they lack the regulatory and resource-allocation frameworks necessary for sustainable adoption, which, ironically, should be informed by the researchers who conduct those works. This disconnect is shaped by a complex array of institutional and economic factors. Researchers often face significant constraints, such as rapid technology lifecycles and rigid academic publication formats, which complicate the integration of deep policy suggestions into relatively short research papers.

### Structural Barriers to Policy Narratives

The significant association between publication venue (journal vs. conference) and the presence of policy implications points to a structural incentive problem in academic publishing. Conference proceedings, which are often the first points of entry for new AI research, may appear to prioritize technical novelty via design-oriented research over policy relevance. This pattern can create a filter where policy thinking is treated as secondary or even peripheral, rather than integral to the study's design. This "policy silence" is likely reinforced by review processes that do not explicitly require authors to define the policy implications of their findings. Our results do not support the rationale that conferences are less feasible for drawing policy implications since their sample size tends to be smaller.

It is not a surprise that, as a very young emerging field, very few papers have enough evidence to specify the detailed conditions for when and where the policy implications are applicable. Our classification suggests that the level of education being suggested is very broad and mainly points to general institutions or general educational settings. Compared to a more mature field of human-resource management research, policy implications can carry the detailed conditions of when and where policy implications may be applicable (Dello Russo et al., 2023). It may take time for AI × SEL research to enrich that level of conditional applications, but researchers should pay more attention to doing research that likely provides insights for policymakers on when and where something works.

**Implications for research, review, and policy**

For researchers, our results describe a framework for writing policy implications that are evidence-grounded and reader-oriented. Researchers can adopt the WH-framework (who, what, why, when/where, how) during their writing process to ensure their conclusions reach beyond the technical domain. As our findings suggest, the lack of policy implications in both quantity and quality urges researchers to rethink how we can proactively create impacts on the real world via our clear, engaging, and thoughtful reader-oriented writing. Given that AI × SEL is inherently practice-oriented, it is important that scientific work in this field does not remain purely abstract but actively considers how findings can inform governance and implementation.

To create impactful implications for the real world, we suggest these actionable writing suggestions:

- Signal policy relevance early: Explicitly state the potential policy impact within the abstract to guide relevant stakeholders to your work.
- Adopt the lexicon of governance: Utilize precise terminology, such as "regulatory framework", "institutional mandate", or "evidence-informed policy", to signal that your conclusions are intended for decision makers.
- Adopt the WH-framework: Apply the "Who, What, Why, When/Where, and How" framework to your implication narratives to enhance clarity, accountability, and feasibility.
- Calibrate rhetorical strength: Align the certainty of your language (e.g., "suggests", "could", "must", "requires") with the strength of the underlying evidence to maintain scholarly credibility and help readers calibrate the strength of evidence-based suggestions.

- Standardize structural placement: Use dedicated, clearly labeled subsections (e.g., "Policy Implications") to ensure findings are easily discoverable for readers.
- Prioritize implementation pathways: Shift the focus from what is needed to how it can be operationalized within existing educational infrastructures.
- Practice stakeholder co-production: Consult relevant stakeholders (educators, administrators, or technologists) during the research and writing phase to ensure proposed policy paths are appropriately grounded.
- Include a "Policy Summary" table: For complex reviews, include a visual aid or table that condenses your policy recommendations into a quick-reference format for busy stakeholders.
- Explicitly address implementation trade-offs: Acknowledge the potential costs, resistance, or conflicts your proposed policy might encounter (e.g., resource constraints, ethical requirements, etc.).
- Delineate actor responsibility: Avoid vague attributions; clearly separate what is expected of national ministries, school boards, university leaders, and educators.
- Address "living" policy: If applicable, emphasize that policy guidelines should be iterative and adaptable, moving away from "static rulebooks" toward flexible, responsive governance frameworks.

For Editors and Reviewers, journal editorial boards should mandate a "Policy and Practice" section in submissions, especially for studies that claim to guide educational stakeholders. We found two papers that contain this mandated section. However, only one paper successfully spelled out policy guidance, while one paper did not provide any implications for policy under its heading of policy implications. Thus, if a paper claims to have policy implications, it should be subjected to scrutiny on its specificity and feasibility.

For Policymakers, rather than waiting for general guidance, educational authorities should actively search for "policy-oriented systematic reviews" that synthesize evidence specifically for their local contexts, rather than relying on disparate, high-level conference papers. Also, as several reviewed papers suggest, collaborating with researchers is essential for policymakers to stay informed of evidence and to translate abstract research findings into context-sensitive, actionable governance ideas that address the complex realities of educational systems and practice.

## 5. Conclusions

The intersection of AI × SEL stands at an intriguing crossroads. While research in this area is exploding, our review reveals a disconnect between technical innovation and governance considerations that may limit the impact of the very innovations we are so eager to implement. After systematically analyzing 65 studies, we observed a troubling pattern: the vast majority of literature focuses on what these tools *can* do, while staying largely silent on the regulatory and institutional frameworks required to use them safely and efficiently. Policy is too often treated as a generic "add-on", a few separated sentences in the middle of a paragraph in the discussion section, rather than an essential component of the scientific writing itself.

We are seeing a "techno-solutionist" trap, where enthusiasm for AI's potential often overshadows the practical realities of the classroom and educational systems. Because academic incentives currently prioritize technical novelty over the slower, more complex work of governance, researchers, reviewers, and editors may not always fully foreground policy considerations. As a result, current publications often provide limited actionable, grounded guidance for teachers and school leaders who need to integrate these tools with confidence.

It is time for a cultural shift in how we write and think about these implications. Moving from "implication-as-afterthought" to "implication-as-methodology" is not only about rearranging a paper or writing a better policy implication section, although that should be taken care of; it is about designing research with policy in mind from day one. By adopting the WH-framework, we can move away from vague, abstract calls for ethical governance and toward the kind of precise, actionable, and accountability-driven narratives that policymakers can actually act upon.

Finally, the future of this field depends on more than just scientific rigor; it depends on the clarity, feasibility, and empathy with which we translate our findings into real-world guidance. We urge our colleagues to move beyond high-level theory or constrained empirical works and start providing a concrete, evidence-based approach to governance that truly reflects the complexity of modern classroom and educational systems.

**Declaration of generative AI and AI-assisted technologies in the manuscript preparation process.**

During the preparation of this work, the authors used Google Gemini to assist with language polishing and with brainstorming keywords for database searches. After using this tool, the authors carefully reviewed and edited the content as needed and take full responsibility for all interpretations and conclusions presented in the article.